\documentclass[]{mn2e}  
\usepackage[dvips]{graphicx}
\usepackage{mnras_cite}  
\usepackage{epsf}
\def\ga{\mathrel{\mathpalette\fun >}}
\def\fun#1#2{\lower3.6pt\vbox{\baselineskip0pt\lineskip.9pt
\ialign{$\mathsurround=0pt#1\hfill##\hfil$\crcr#2\crcr\sim\crcr}}}

\def\rn{} 
\def\nnn#1#2 #3{#2. #3. #1}                    
\def\nnnn#1 #2 #3 #4{#2. #3. #4. #1}            
\def\nnnnn#1 #2 #3 #4 #5{#2. #3. #4 #5. #1}     
  
\def\rf#1;#2;#3;#4;#5 {{\frenchspacing\par\rn#1, #3 {\bf #4}, #5 (#2). \par}} 
\def\rfbook#1;#2;#3;#4;#5 {{\frenchspacing\par\rn#1, {\it #3} (#5, #4,#2).\par}} 
\def\rfprep#1;#2;#3 {{\par\frenchspacing\rn#1, #3 (#2).\par}}

\newcommand{\nc}{\newcommand}

\nc{\be}{\begin{equation}} \nc{\ee}{\end{equation}}
\nc{\beq}[1]{\begin{equation}\label{#1}} \nc{\eeq}{\end{equation}}
\nc{\bea}[1]{\begin{eqnarray}\label{#1}} \nc{\eea}{\end{eqnarray}}

\nc{\inv}[1]{\frac{1}{#1}}
\nc{\pa}{\partial} \nc{\na}{\nabla}

\nc{\ul}{\underline} \nc{\al}{\alpha} \nc{\g}{\gamma}
\nc{\Del}{\Delta} \nc{\e}{\textrm{e}} \nc{\eps}{\epsilon}
\nc{\lam}{\lambda} \nc{\Om}{\Omega} \nc{\Omm}{\Omega_m}
\nc{\Oml}{\Omega_\Lambda} \nc{\LCDM}{$\Lambda$CDM~} 
\nc{\ve}{\varepsilon} \nc{\mn}{{\mu\nu}} \nc{\vp}{\varphi}

\def\gsim{\; \raise0.3ex\hbox{$>$\kern-0.75em
\raise-1.1ex\hbox{$\sim$}}\; }

\nc{\de}{\delta} 
\nc{\tISW}{\triangle_T^{ISW}}
\nc{\hn}{\hat{n}}
\nc{\bH}{\bar{H}} 
\nc{\Ol}{\Om_{\Lambda}} 
\nc{\w}{\omega} 

\begin{document}


\title[New light on Dark Cosmos]{New light on Dark Cosmos}


\author[Gazta\~{n}aga, Manera \& Multam\"aki]{Enrique Gazta\~naga$^{1,2}$, 
Marc Manera$^1$, Tuomas Multam\"aki$^3$ \\
$^{1}$ Institut d'Estudis Espacials de Catalunya, IEEC/CSIC, 
F. de Ciencies, Torre C5 Par 2a, UAB, Bellaterra (08193 BARCELONA)
\\
$^{2}$ Visiting: INAOE, Tonanzintla, P.O.Box 51, 7200 Puebla, Mexico
\\
$^{3}$ NORDITA, Blegdamsvej 17, DK-2100, Copenhagen, Denmark
}
 
 

\maketitle  

\begin{abstract} 
Recent studies by a number of independent collaborations, have correlated
the CMB temperatures measured by the WMAP satellite with different galaxy
surveys that trace the matter distribution with light from the whole range
of the electromagnetic spectrum: radio, far-infrared, optical and X-ray
surveys. The new data systematically finds positive correlations,
indicating a rapid slow down in the growth of structure in the universe.
Individual cross-correlation measurements are of low significance, but we
show that combining data at different redshifts introduces important new
constraints. Contrary to what happens at low redshifts, for a fixed $\Omm$,
 the higher the dark energy contend, $\Ol$, the lower the ISW
cross-correlation amplitude. At 68\%  confidence level, the data finds new
independent evidence of dark energy: $\Ol =0.42-1.22$ .  It also confirms,
to higher significance, the presence of a large dark matter component:
$\Omm =0.18-0.34$, exceeding the density of baryonic matter, but far from
the critical value.  Combining these new constraints with the prior of a
flat universe, or the prior of an accelerating universe  provides strong
new evidence for a dark cosmos. Combination with supernova data yields $\Ol
= 0.71 \pm 0.13$, $\Omm = 0.29 \pm 0.04$. If we also assume a flat
universe,  we find $\Ol = 0.70 \pm 0.05$ and $w = -1.02 \pm 0.17$ for a
constant dark energy  equation of state. 
\end{abstract} 



\maketitle


\section{Introduction}

In the last few years a new cosmological scenario with a significant
smooth Dark Energy (DE) component has emerged. The Cosmic Concordance Model
(CCM, from now on) is a spatially flat universe with baryons ($\Om_b \sim
4\%$), cold dark matter ($\Omega_{CDM} \sim 23\%$) and a significant DE
component ($\Oml \sim 73\%$). The model is well supported by the supernova
type Ia observations (SNIA) \cite{riess,perlmutter}, observations on
large scale structure (LSS) \cite{sdss,2df} and the cosmic microwave
background experiments (CMB), in particular by the recent WMAP
experiment \cite{wmap}. The energy density of the universe seems
dominated by the unknown DE component, presenting a
formidable observational and theoretical challenge.
The three key observational probes measure complementary aspects of
the cosmological parameter space. The SNIA indicate that the universe
is accelerating but present data is degenerate  
for alternative cosmological scenarios. The LSS
observations constrain $\Omm$ but leave the DE question
unanswered. Constraints from primary anisotropies in the  CMB
indicate that we live in a flat universe 
but require a prior on the value of the local
Hubble rate $H_0$\cite{Blanchard}. 
Assuming that the universe is
well described by a $\Lambda$CDM model, combining all these three
observations gives us the cosmological CCM model.

The Integrated Sachs Wolfe effect, ISW, \cite{ISW} is a direct probe
for the (linear) rate of structure formation in the universe.
Secondary anisotropies in the CMB appear because of the net
gravitational redshifts affecting CMB photons that travel through an
evolving gravitational potential $\Phi$. These secondary temperature
anisotropies are therefore correlated with local, evolving, structures
on large scales. The correlation is negative when structures grow, as
increasing potential leaves a cold spot in the CMB sky, and positive
otherwise. In a flat universe without DE (Einstein-deSitter,
or EdS, model) this cross-correlation is expected to be zero because
the gravitational potential remains constant, despite the linear
growth of the matter fluctuations.

The rate of structure formation in the universe can also be measured
by galaxy peculiar velocities or galaxy redshift distortions, on very
large scales through the so-call $\beta$ parameter determination
\cite{peacock,pope}. The ISW effect provides an independent and
complementary probe of the same effect. Independent, because it uses
temperature anisotropies instead of the velocity field, and
complementary, because of the different assumptions and systematics
that relate measurements with theory. 
Despite  recent advances in the
size of galaxy redshift surveys such as SDSS and 2dFGRS, the spectrum
of matter fluctuations $P(k) \propto <\delta(k)^2>$  is quite
difficult to measure directly over very large scales
\cite{sdss,2df,APMPk}. Part of the problem is that matter correlations
fall quickly to zero on scales larger than $30$ Mpc/h ($k <0.1$
h/Mpc). In contrast, fluctuations in the gravitational potential go as
$\Phi(k) \propto \delta(k)/k^2$ and therefore  extend  over larger
distances, which makes the signal more detectable (see also comments
to Fig. \ref{LCDMgraf}). The ISW cross-correlation traces the
gravitational potential, $\Phi$, and thus provides a new window
to study the largest structures,
extending over several degrees in the
sky or tens of Mpc/h at the survey depth.

\section{Growth of density perturbations}

Gravitational evolution of matter fluctuations,
$\delta={\rho/\bar{\rho}}-1$, is dependent on the cosmological model
via the evolution of the scale factor $a=a(t)$. Compared to a static
background, a rapidly expanding background will slow down the collapse
of an over dense region. In the linear regime, a small initial
perturbation $\delta_0$ grows according to the growth factor $D(t)$:
\beq{D(t)}
\delta(t)= D(t) ~\delta_0
\eeq
which, under quite generic assumptions, 
eg \cite{lobo,Multamaki1,lue}, follows a simple
harmonic equation: 
\beq{lineareq} {d^2 D\over{d\eta^2}}+\left(2+{\dot{H}\over
H^2}\right) {dD\over{d\eta}} +3 c_1~D=0, 
\eeq 
where  $\eta=\ln(a)$ is the conformal
time and $H=H(\bar{\rho}) \equiv \dot{a}/a$ is the background Hubble
rate ($\dot{a}$ and $\dot{H}$ are proper time derivatives). 
For a flat cosmological model with a generic dark energy
equation of state 
\beq{eq:w(z)}
p=w(z)\rho
\eeq
 we then have: 

\begin{equation}
H^2=\left({\dot{a}\over a}\right)^2 = H_0^2 \left[ \Omm (1+z)^3+\Oml
e^{3\int_0^{z} {dz'\over 1+z'}(1+w(z'))}  \right]
\label{friedman}
\end{equation}

where $\Oml$ and $\Omm$ are the dark
energy and dark matter densities today in units of the critical
density $\rho_c \equiv  3~H^2/(8\pi G)$. And $c_1$ is given by:

\begin{equation}
c_1=-{1\over 2} {H^2_0 \Omm (1+z)^3 \over H^2(z)}
\end{equation}


In this paper we study two cases. A generic (not necessarily flat) $\Lambda$CDM 
model where DE density is constant over the evolution of the universe $(w=-1)$; and
a flat $\Lambda$CDM model with a constant equation of state parameter. For those
models we have

\begin{equation}
c_1  =-(1/2)~\Omm/(\Omm+\Oml a^{3})
\end{equation}


One may choose to compare the results to the EdS model: $\Oml=0,\ \Omm=1$, 
in which case the solution to Eq. (\ref{lineareq}) is $D\propto a$. 
This means that $\delta$ grows linearly with the
scale factor, $\delta \propto a$, while the  corresponding
gravitational potential fluctuation, $\Phi \sim \delta/a$, remains
constant as the universe expands. For non EdS models $\Phi$ would
change during the expansion of the universe which would turn into 
a galaxy-CMB temperature cross-correlation signal.

\subsection{The ISW effect}

ISW temperature anisotropies are given by \cite{ISW}: 
\beq{ISW1} 
\tISW(\hn) \equiv {T(\hn)-T_0\over{T_0}} =- 2 \int dz ~{d\Phi\over{dz}}(\hn,z)
\eeq 
where $\Phi$ is the Newtonian gravitational potential at redshift
$z$.  One way to detect the ISW effect is to cross-correlate
temperature fluctuations with galaxy density fluctuations projected in
the sky \cite{CrTu96}. On large linear scales
and small angular separations, the
cross-correlation $w^{ISW}_{TG}(\theta)  =  <\tISW(\hn_1)
\de_G(\hn_2)>$ is \cite{Fosalba1}: 

\bea{final_wtg}
w^{ISW}_{TG}(\theta) & =  & 
{1\over{2\pi}}\int {dk\over k} P(k) g(k\theta) \nonumber \\ 
g(k\theta) & = & \int dz\, W_{ISW}(z) W_G(z) {H(z)\over c}
J_0(k\,r_A\theta)  \nonumber\\ 
W_{ISW}(z) & = & 3 \Omm (H_0/c)^2  {d[D(z)/a]\over dz} \\ 
W_G(z) & = &  b(z)\phi_G(z) D(z),\nonumber 
\eea

where $J_0$ is the zero order Bessel function, $\phi_G$ is the survey
galaxy selection function along the line of sight $z$ and $r_A=r_A(z)$
the comoving transverse distance. 
The power spectrum is
$P(k)= A~k^{n_s}~T^2(k)$, where $n_s\simeq 1$\footnote{
Throughout the paper we made the assumption
of scale invariant primordial fluctuations ($n_s\simeq 1$).
For other possibilities see, eg, \cite{Subir}.}
and $T(k)$ is the \LCDM transfer function, which we evaluate using
the fitting formulae of Einseintein \& Hu 1998.
We make the common assumption that galaxy and
 matter fluctuations are related through the linear
bias factor, $\de_G(\hn,z)=b(z)\de_m(\hn,z)$.  

\begin{figure}
{\epsfxsize 3.0in  
\includegraphics[width=70mm,angle=-90]{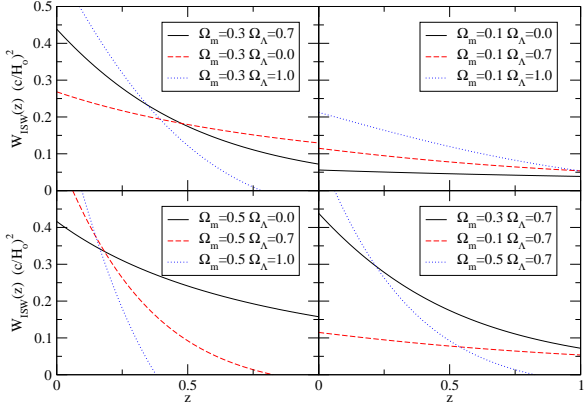} 
\vskip -0.3in}
\caption{Redshift dependence of $W_{ISW}(z)$ in Eq.[\ref{final_wtg}] for different values of $\Omm$ and
$\Oml$.
Bottom left, top right and top left  panels shows a fixed  $\Omm=0.5$,
$\Omega_m=0.3$ and $\Omega_m=0.1$ respectively. In all 
cases: $\Omega_\Lambda=0.0$ (dotted blue line) , $\Omega_\Lambda=0.7$ (continuos black line) and $\Omega_\Lambda=1.0$ (dashed red line).
Bottom right panel shows a fixed $\Omega_\Lambda=0.7$ and 
$\Omega_m=0.3$ (continuos black line), $\Omega_m=0.5$ 
(dotted blue line) and $\Omega_\Lambda=0.1$ (dashed red line).}
\label{ISWfig} 
\end{figure}

For the $\Lambda$CDM case
the ISW effect is non-zero, and the kernel $W_{ISW}$ 
can be well approximated by
$W_{ISW}(z)= - 3 \Omm (H_0/c)^2 D(z) (f-1)$, where f is the relative
growth factor, $f\simeq \Omm(z)^{6/11}$. 
$W_{ISW}$ decreases as a function of increasing redshift
and goes to zero both for $\Omm  \rightarrow 0$ 
and for $\Omm  \rightarrow 1$. At low redshifts,
the ISW effect is larger for larger values of $\Omm$,
but the redshift evolution depends on the curvature (ie how
quickly the $H$ and $D$ evolve to the EdS case). This is illustrated in
Fig.\ref{ISWfig} which shows how $W_{ISW}$ depends on $z$ for
different values of $\Oml$ and $\Omm$. At high redshifts,
the lower the value of $\Oml$ (for a fixed $\Omm$) 
the larger the ISW amplitude.

\begin{figure}
{\includegraphics[width=70mm,angle=-90]{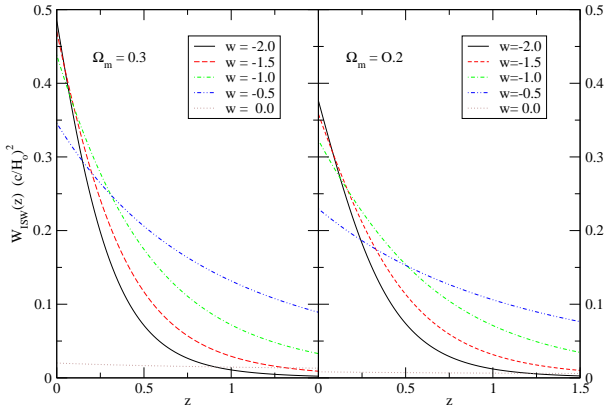}}
\caption{Redshift dependence of $W_{ISW}$ in eq [\ref{final_wtg}] for
flat models with constant equation of state. Right panel shows a fixed
$\Omm=0.2$ and left panel $\Omm=0.3$. In both cases $w=-2$ (black continuous
line), $w=-1.5$ (red dashed line), $w=-1$ (green dot-dashed line), $w=-0.5$
(blue doubledot-dashed line), and $w=0$ (brown dotted line)}
\label{WISWtwo}
\end{figure}

\begin{figure}
{\includegraphics[width=70mm,angle=-90]{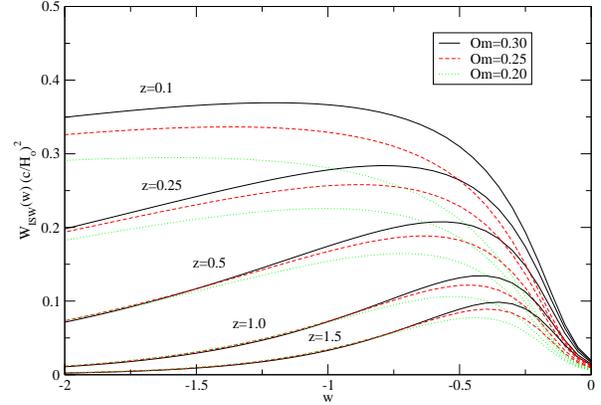}}
\caption{Dependence of $W_{ISW}$ on the equation of state parameter $w$
for flat models at different redshifts and values of $\Omm$. $\Om=0.3$ (back
continuous lines), $\Omm=0.25$ (red dashed lines), $\Omm=0.2$ (green dotted
lines). Redshifts are 0.1, 0.25, 0.5, 1.0, 1.5 from top to bottom.}
\label{varyingw}
\end{figure}

In Figures \ref{WISWtwo} and \ref{varyingw} we also shown for a given 
flat cosmology model the dependence of the $W_{ISW}$ on redshift and
on the equation of state parameter w. For a given redshift and $\Omm$ 
there exists a maximum of $W_{ISW}$ 
around $w=-0.5$. This maximum would translate into a maximum in the 
cross-correlation signal $w_{TG}$. If data turns out to be greater than
this maximum this would clearly disfavor models with constant equation
of state.

\subsection{Bias Self-calibration}
\label{sec:bias}
  
Linear bias is used to study how well light
traces the  underlying statistics of linear  matter
fluctuations. 
On these very large scales, fluctuations $\delta$ 
are small and linear theory works very well both for biasing and gravity. 
We remove the effects of biasing 
in our parameter estimation by comparing the
observed galaxy-galaxy correlation $w_{GG}$, in the very same
samples used for the cross-correlation,
to the matter-matter
correlation $w_{mm}$ predicted by each model \cite{Fosalba2}. 
The effects of bias are also redshift dependent, but
given a galaxy selection function $\phi_G(z)$, picked at $z=\bar{z}$,
we approximate the bias with a constant $b=b(\bar{z})$ for that
particular survey. We
then have: $w_{TG}  = b(\bar{z}) w_{Tm}$ and  $w_{GG} = b^2(\bar{z})
w_{mm}$, so that an effective  linear bias $b$ can be estimated as the square
root of the ratio of galaxy-galaxy and matter-matter correlation
functions: 
\beq{wbias} 
b  =  \sqrt{{w_{GG}\over w_{mm}}}. 
\eeq 
Such prescription has been shown  to work well in a variety of
galaxy models (eg see \cite{berlind}). The
values of $w_{mm}$ can be computed similar to (\ref{final_wtg}) by

\bea{final_wmm}
w_{mm}(\theta) & =  &
{1\over{2\pi}}\int dk k  P(k) g(k\theta) \nonumber \\
g(k\theta) & = & \int dz\, W_m^2(z) {H(z)\over c}
J_0(k\,r_A\theta)  \\
W_m(z) & = & \phi_G(z) D(z),\nonumber
\eea
where the only difference between $W_m$ and $W_G$ is the 
bias factor $b(z)$ in Eq.[\ref{final_wtg}].
Note how the estimation of $b$ in Eq.[\ref{wbias}] depends on the normalization 
of the power spectrum in $w_{mm}$. We choose to normalize each model by fixing 
$\sigma_8$. To make our results independent of this normalization
we will marginalize over $\sigma_8$ and $h$. Taking flat priors
and ranges $\sigma_8=0.8-1.0$ and $h=0.72-0.77$.
We compare the predictions with the observational data $w_{TG}$ 
normalized to the
CCM model bias, ie $w_{TG}/b$, where $b$  is estimated from Eq.
(\ref{wbias}) using $w_{mm}$ in the CCM model.
Consequently, for other models, we will need to renormalize each of the theoretical 
predictions to the CCM model bias using a ``relative bias'':
$w_{TG}^{mod}/b=b_r  \; w_{tm}^{mod}$, where
$b_r^2= {w_{mm}/w_{mm}^{mod}}$ is the ratio of the concordance model
prediction to the one in the corresponding model. We choose to
estimate this relative bias at $R=8$ Mpc/h, but the actual number has little
effect in our final conclusions.

\section{Observational Data} 

Recent analysis by independent collaborations, have cross-correlated the
CMB anisotropies measured by WMAP with different galaxy surveys. The 
median galaxy redshifts expand over a decade (ie $0.1<\bar{z}<1.0$) and
trace the matter distribution with light from the whole range of the
electromagnetic spectrum: radio, far-infrared, optical 
and X-ray surveys (see Table \ref{taula}). 
The cross-correlation and error estimation techniques used are also quite 
different but they yield comparable results over the scales of interest.
Compare for example the  Montecarlo errors to jackknife errors 
in Fig.3 in \scite{Fosalba1}.
In our compilation of the different data sets, we average the results
on fixed angular scales around $\theta=6^\circ$.  This corresponds to
proper distances of $\simeq 25$Mpc/h at $\bar{z} \simeq 0.1$ and
$\simeq 100$ Mpc/h at $\bar{z} \simeq 1.0$ in the CCM model and avoids
possible contamination from the small scale SZ and lensing effects, eg
see Fig. 3 in \scite{Fosalba2}.

Radio galaxies from NVSS \cite{nvss} and hard X-ray background
observed by HEAO-1 \cite{xray}, have been cross-correlated with WMAP
data \cite{Boughn_Nat,Boughn_A}, to find a signal of $1.13\pm 0.35\,$ 
times the CCM model prediction at $z\sim 0.9$. The different biases for
X-rays, $b^2=1.12$, \cite{BoughnXray} and for radio galaxies,
$b=1.3-1.7$, \cite{BoughnRad} have been taken into account. A
compatible signal has also been found with the NVSS data by the WMAP
team \cite{Nolta}. 

The cross correlation of WMAP 
with  galaxies ($17<b_J<20$) in the APM Galaxy Survey
\cite{APM} (covering about 20\% of the South Galactic Cap, SGC) 
was found to be $w_{TG}=0.35\pm 0.13\, \mu K$  
at scales  $\theta=4-10^\circ$  with  $b
\simeq 1$ \cite{Fosalba1}. The cross-correlation of WMAP with the SDSS DR1
\cite{sdss} (covering about 10\% of the
North Galactic Cap, NGC) have been done for several subsamples
\cite{Fosalba2}. The first
sample  ($\bar{z}\sim 0.3$) contains $\sim 5$ million objects 
classified as galaxies in SDSS (with $ r < 21 $ and low associated
error).  For this sample, which has  $b \simeq 1$,
$w_{TG}=0.26\pm0.13\, \mu K$ at scales $\theta=4-10^\circ$. The
high redshift sample ($z\sim 0.5$) has
$w_{TG}=0.53\pm0.21\, \mu K$ and $b^2\simeq 6$. 
The SDSS data has also been cross-correlated with
WMAP by the SDSS team \cite{Scranton} using nearly 25 million galaxies
in four redshift samples. Their results are similar with
those obtained earlier by \scite{Fosalba2} but no bias from galaxy-galaxy auto
correlation function is given. The infrared 2MASS Galaxy Survey
\cite{2mass},  with  $z\sim 0.1$, show a WMAP cross-correlation 
of $1.53\pm 0.61$ times the CCM prediction, 
with a bias of $b=1.18$ \cite{Afshordi}.

We have selected independent measurements for which
the bias CCM $b$ (from $w_{GG}$)
is known, so that we can applied the bias ``self-calibration'' proposed
in section \S\ref{sec:bias}. The data is summarized in Table \ref{taula} 
and displayed in Fig.1. In the results below we also include a $10\%$
uncertainty in the median redshift. We chose the  values of NVSS+HEAO-1
quoted by \cite{Boughn_A} as representative of both the \scite{Nolta}
and \scite{Boughn_Nat} analysis. 
For the SDSS, we chose the values in
\scite{Fosalba1} where the CCM bias $b$ is estimated using
Eq.(\ref{wbias}). Note how the selected samples are complementary.
The samples which have large sky overlap (eg 2MASS and NVSS+HEAO-1)
have negligible redshift overlap.
When the redshift overlap is significant (ie in 2MASS-APM or SDSS-NVSS
could be up to $20\%$) the sky overlap is small (less than $10\%$).
Consequently, the different samples in Table \ref{taula}  have less than 
$1\%$ volume in common. This is negligible,  given that individual sampling errors (which are proportional to volume) are of the order of 30\%.

The most significant detection in Table \ref{taula}
seems to be the one quoted by \scite{Boughn_A}
for the NVSS+HEAO-1 samples. Given the systematic 
uncertainties involved in the bias and selection function of both of these samples, we have checked that our results do not
changed much (less than $20\%$ in the area of the contours in Fig.3)
when we double the quoted errorbar. 
Doubling this errorbar corresponds to an additional $50\%$ systematic 
uncertainty in the value $b$ or to a $40\%$ uncertainty in the
median redshift of the samples.

The observational
data not included in Table \ref{taula} is in good agreement with the
values in the table, but is excluded to avoid redundancy. The
agreement of the redundant data provides further confirmation 
and indicates that errors are dominated by sampling variance
rather than by the methodology or the systematics.

\begin{table}
\begin{tabular}{|c|c|c|c|} \hline
$\bar{z}$ & $w_{TG}/b$ & $b$ & catalog,  Band \\ \hline 0.1   & $0.70
\pm 0.32$  & 1.1 & 2MASS,  infrared ($2\mu m$) \\ 0.15  & $0.35  \pm
0.17$  & 1.0    & APM,  optical ($b_j$) \\ 0.3   & $0.26  \pm 0.14$  &
1.0    & SDSS, optical ($r$)\\ 0.5   & $0.216 \pm 0.096$ & 2.4  & SDSS
high-z, optical ($r$+colors) \\ 0.9   & $0.043 \pm 0.015$ & 1-2  &
NVSS+HEAO, Radio \& X-rays\\ \hline \end{tabular} \caption{
Observed  cross correlation $w_{TG}/b$ (averaged for
$\theta \simeq 4-10^\circ$.) of WMAP anisotropies with different
catalogs. Errors in $w_{TG}/b$ includes
20\% uncertainty in $b$. Errors in the median redshift $\bar{z}$ are 
about 10\% .}
\label{taula} 
\end{table}

\begin{figure}
\epsfxsize 2.7in 
\epsfbox{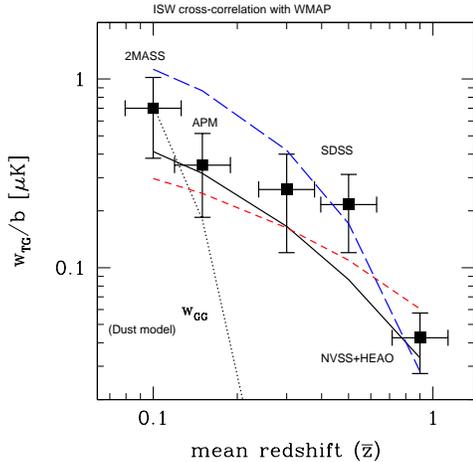} 
\vskip -0.3in
\caption{Symbols with error bars correspond to the different
measurements $w_{TG}/b$ in Table 1. As an illustration of the shape,
the continuous, short-dashed 
and long-dashed lines show the concordance 
($\Omega_m=0.3,\Omega_\Lambda=0.7$), opened  
($\Omega_m=0.3,\Omega_\Lambda=0.0$) and closed ($\Omega_m=0.3,\Omega_\Lambda=1.1$)  
model predictions
 (at $\theta=6^\circ$). The dotted line corresponds to the
galaxy-galaxy prediction (and also the dust contamination model). All lines have
arbitrary normalization.}
\label{LCDMgraf}
\end{figure}

\section{Results}

Fig.\ref{LCDMgraf} compares the $w_{TG}$  observations
with predictions
for a fixed value of $\Omega_m=0.3$ and three different
values of $\Omega_\Lambda$.
We can see how the shape of the prediction depends on the
amount of dark energy. Even though $W_{ISW}$ at $z=0$ depends only
weakly on  $\Omega_\Lambda$, the evolution with redshift
depends more strongly on   $\Omega_\Lambda$. For a fixed $\Omega_m$,
models with larger values of $\Omega_\Lambda$ evolve more rapidly 
with redshift to the EdS case, where the ISW effect vanishes.
Thus, contrary to what happens at $z=0$, the lower the value of $\Omega_\Lambda$ (for a fixed $\Omega_m$) the larger the ISW amplitude at high redshifts
(see also Fig.\ref{ISWfig}).

\begin{figure}
\epsfxsize 2.5in \epsfbox{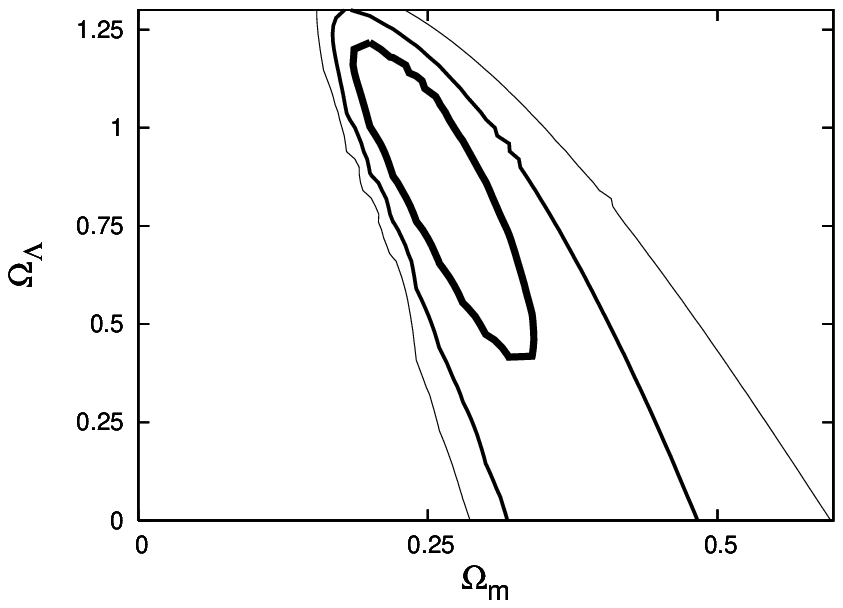} 
\epsfxsize 2.5in \epsfbox{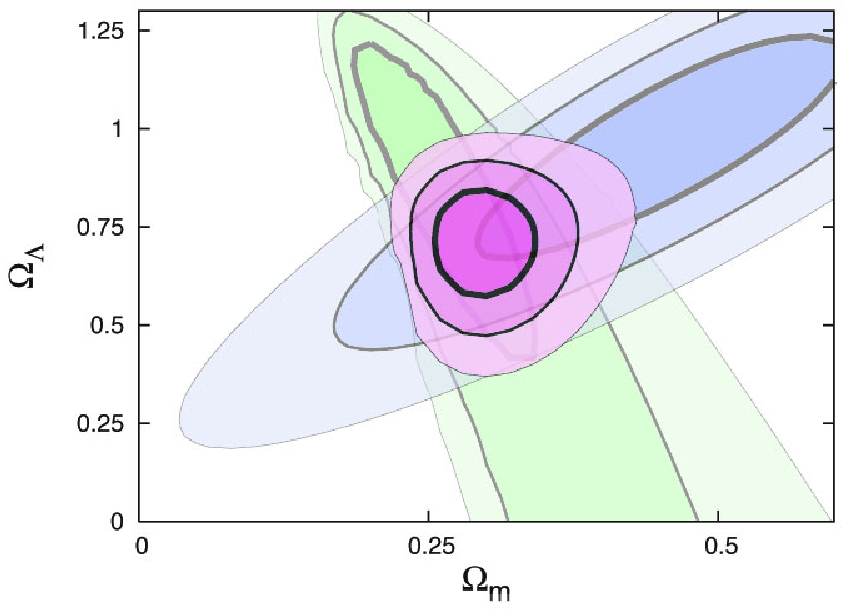} 
\caption{One, two and three sigma confidence
contours in the $(\Omm,\Ol)$ plane (marginalized over $h$) for the
$\Lambda$CDM model. Top: constraints from only ISW. Bottom:
constraints from SNIa (blue) and ISW (green) along with the combined
contours (purple).} \label{combinedconts} 
\end{figure}

To test model predictions with the data, we use a standard
$\chi^2$-test, $\chi^2 = \sum_i ~(O_i -T_i)^2/\sigma_i^2$, where $O_i$
and $\sigma_i$ correspond to the different measurements and errors
and $T_i$ correspond to the
model. The label $i$ runs for $i=1$ to $i=5$ marking the different
data points (column 1 in Table \ref{taula}) as we move in redshift.
In order to take into account the error in the median redshift we take:
\beq{eq:sigmaz}
\sigma_i^2=\sigma_w^2+\left({{d (w_{tg}/b)}\over{dz}}\right)^2\;\sigma_z^2
\eeq
where $\sigma_z$ and
$\sigma_w$ are the errors in the $w_{TG}/b$ and $\bar{z}$ respectively
(see Table \ref{taula}).
We use the relative $\chi^2$ values,
$\chi^2-\chi^2_{min}$, to define confidence levels in parameter
estimation. Top panel of Fig.\ref{combinedconts} shows the 
resulting confidence contours.
 Taking $T_i=0$ we evaluate the significance of the
combined ISW detection. We find that this null hypothesis is rejected
with a very high probability: $P \simeq 99.997\%$  
(from $P_{\nu=4}(\chi^2>26)\simeq 3 \times 10^{-5}$).
We next compute the expected ISW effect and compare
it with the observational data  within the $\Lambda$CDM
family of models, where $\Omm$, $\Om_\Lambda$ and $h$
\footnote{We use $H_0 \equiv 100~h$ km/s/Mpc.}  are
free parameters (we fix the baryonic content $\Omega_b \simeq 0.05$ and the
primordial spectral index $n_s \simeq 1$). 
We choose to normalize each model by fixing $\sigma_8$.
To make our results independent of this normalization we will marginalize
$\sigma_8$ over the range $\sigma_8=0.8-1.0$ (flat prior used).
As we compare $w_{GT}$
normalized to the CCM model bias, we need to compute the relative bias
for other LCDM models (see section \S\ref{sec:bias}).
We choose to estimate this at $R=8 Mpc/h$, but
the actual number has little effect in the conclusions. We have also
marginalized over $h$ in a flat prior range $h=0.72-0.77)$. Our results
are not very sensitive to the ranges used for $\sigma_8$ and $h$: increasing
these ranges by a factor of two change our contours in less than 20 \%.

The best fit using only ISW data corresponds to $\Omm\simeq 0.26 \pm 0.08$, $\Oml \simeq
0.82 \pm 0.40$, in good agreement with other cosmological probes
mentioned above. Bottom panel of  Fig. ref{combinedconts} we show the confidence
contours for a $\Lambda$CDM model along with the constraints from
recent SNIA \cite{barris} observations. From the figure it is clear
how the ISW effects gives new complementary information about the
cosmological parameters. The EdS model is ruled out to high
significance. The confidence contours are almost perpendicular to the
SNIA contours, allowing to constrain the parameter space of the
model well with just these two observations.  Combination of ISW
with supernova data yields $\Ol = 0.71 \pm 0.13$ and 
$\Omm = 0.29 \pm 0.04$. 

\subsection{Uncertainties in the selection function}

We explore here how robust are our results
to the uncertainties in the galaxy selection function.
We take a generic parametric form of the type:

\beq{selection}
\phi(z)dz={1\over \Gamma({m+1\over \beta})} \beta {z^m \over z_0^{m+1}} 
e^{\left(-{z\over z_0}\right)^\beta} dz
\eeq

so that it is normalized to unity. Parameters $\beta$ and $m$  
control the shape of the function and are treat as fix parameters;
$z_0$ is being changed accordingly to the median redshift $\bar{z}$ 
we want for the selection function. When computing our results we 
use $\beta=1.5$ and $m=2$, in which case $\bar{z}=1.41{z_0}$. 

\begin{figure}
\includegraphics[width=60mm,angle=-90]{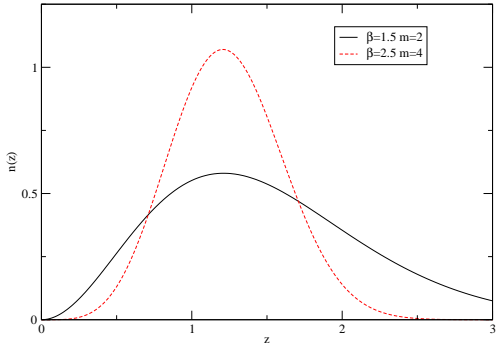}
\caption{Two different selection functions with the same median redshift $\bar{z}=1.41$. 
Both have the generic form given by the equation [\ref{selection}]. The black
continuous line corresponds to $\beta=1.5$ and $m=2$ while the red dashed line
is for $\beta=2.5$,$m=4$}
\label{fig:beta}
\end{figure}

\begin{figure}
{\epsfxsize 2.5in \epsfbox{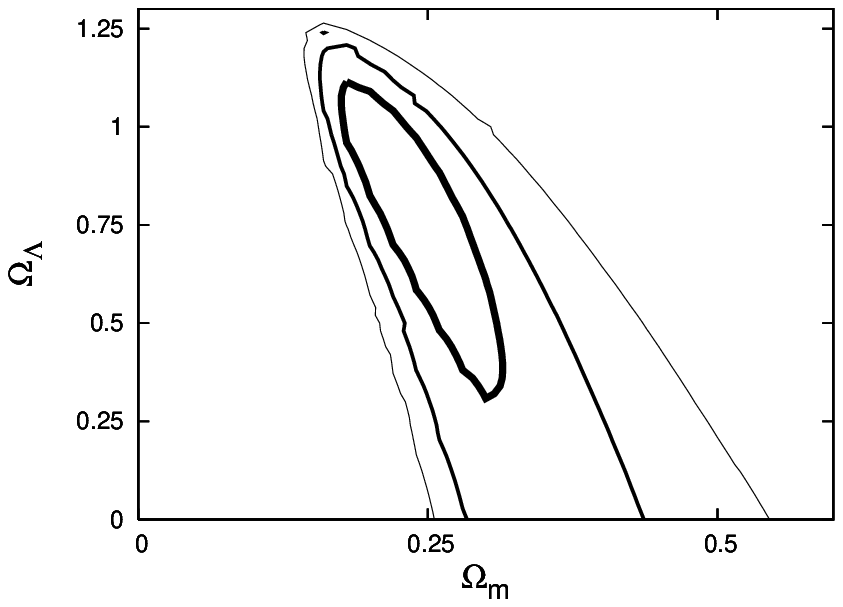}}
{\epsfxsize 2.5in \epsfbox{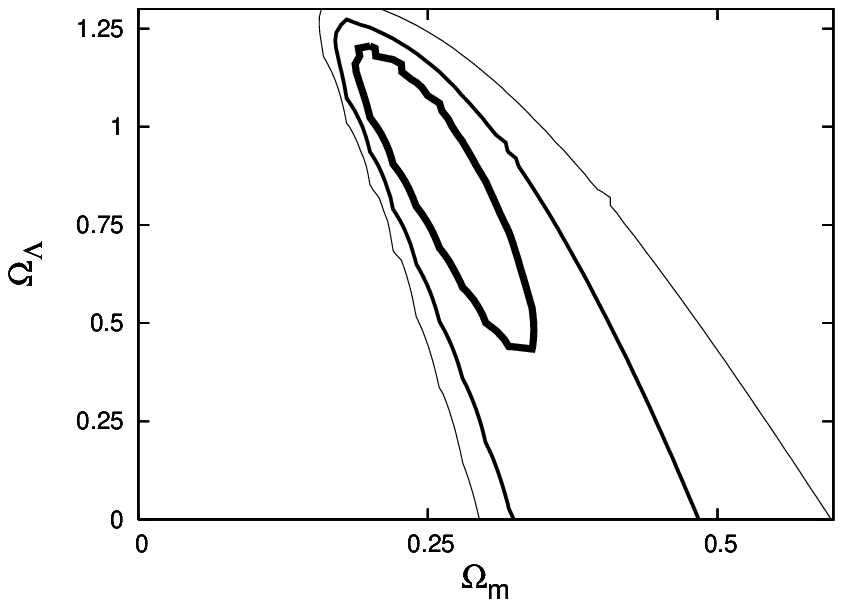}}
\caption{One, two and three sigma confidence contours in the $(\Omm,\Oml)$
plane for the $\Lambda$CDM model. Top panel:  contours using  a more peaked 
selection function ($\beta=2.5$,  $m=4$) but with the same mean redshift as the fiducial case
(ie compare to Fig.\ref{combinedconts}).
Bottom panel: contours when errors in the median 
redshift of the selection functions are neglected.
}
\label{selectionfunc}
\end{figure}

In order to clarify the role of the selection function shape we 
recalculate our results  with a much more peaked selection function.
This second selection function have $\beta=2.5$ and $m=4$ and it
is plotted together with the fiducial one in Figure \ref{fig:beta}. Both
cases have the same median redshift $\bar{z}=1.41$. 
Top panel of Fig.\ref{selectionfunc}
shows the contours in the $(\Omm,\Oml)$ plane  
for the more peaked selection function (with $\beta=2.5$ and $m=4$).  
The contours are similar to the 
fiducial model (ie compare to Fig. \ref{combinedconts}) but favoring 
slightly lower values for $\Oml$ and $\Omm$.

Besides the uncertainty on the shape of the selection function \
there is also uncertainty in the median redshift. We have
checked what happens if this uncertainty is not taken into account. 
We just set the redshift errors $\sigma_z=0$ in Eq.[\ref{eq:sigmaz}].
Contours for the $(\Omm,\Oml)$ plane are plot in the 
bottom panel of figure \ref{selectionfunc}, 
which are also to compare with figure \ref{combinedconts}. 
There is hardly any difference because the theoretical values of
 $w_{TG}$ change very little within
the median redshift error range.

\subsection{Equation of state}

\begin{figure}
\epsfxsize 2.5in \epsfbox{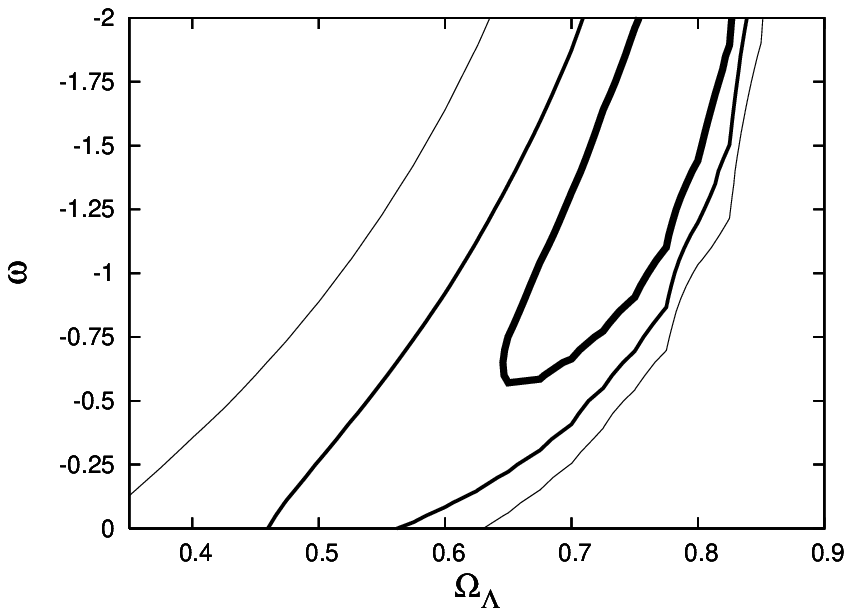}
\epsfxsize 2.5in \epsfbox{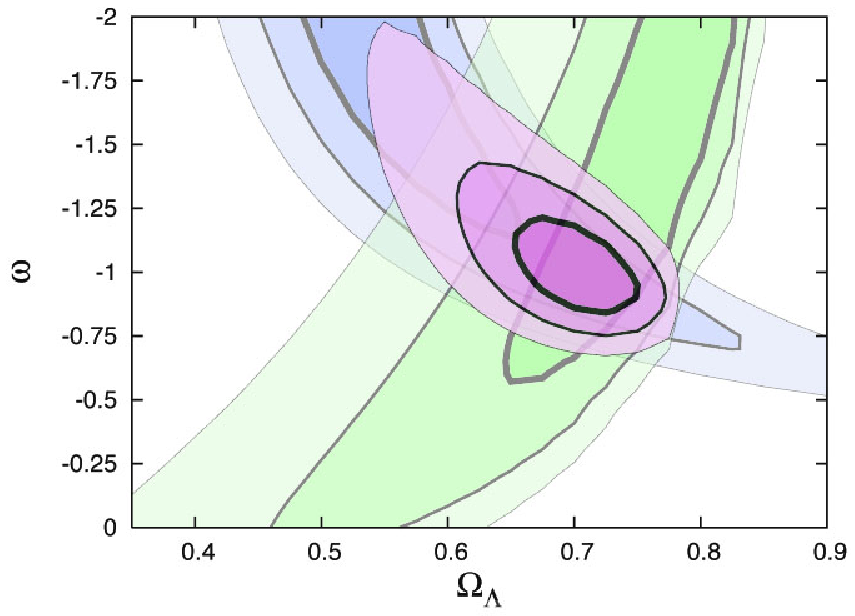}
\caption{One, two and three sigma confidence
contours in the $(\Oml,w)$ plane (marginalized over $h$ and $\sigma_8$. Top: 
constraints only from ISWfrom ISW. Bottom: constraints from SNIa (blue) and
ISW (green) along with the combination (purple).} 
\label{plotwOm}
\end{figure}

The ISW effect can also be used to constrain the dark energy equation
of state parameter. In this case, as suggested by the CCM, we assumed
a flat universe. We focus on a constant $w$ parameter and 
maintain the same flat priors for $h$ and $\sigma_8$ 
( $0.72 <h< 0.77 \;\; 0.8 <\sigma_8 < 1.0$ ). Top panel of Figure \ref{plotwOm}
show the one,two and three sigma contours for the ($\Omm,w$) plane using 
only the ISW data. Join contours with the SNIa data are shown in the
bottom panel of  Figure \ref{plotwOm}. Both datasets are also complementary
for the $w$ determination. The SNIa data is from \cite{barris}.

Making a join ISW+SNIa analysis with the flat prior reduces notably the allowed space for
the parameters to $w = -1.02 \pm 0.17$ and $\Ol = 0.70 \pm 0.05$.
The contours are comparable with other analysis in 
literature \cite{Tegmark2} which combines SNIa with WMAP and SDSS data.
The results we found are still in full agreement to the 
CCM with $\Oml\simeq 0.7 $ and $w = -1$.

\subsection{Possible Contaminants}

The constraining power of the new ISW data comes from the simultaneous
fitting of data at different redshifts, that is from the shape
information in Fig. \ref{LCDMgraf}. Because of the uncertainties in
the relative normalization due to a relative bias, any given point
alone does not constrain well the cosmological parameters. 
But the combination of the data gives us a new powerful tool for cosmological
parameter estimation. 

The shape of the curve as a function of redshift
also provides an important test for systematics.
CMB and galaxy maps are both masked and corrected from galactic
absorption/extinction, but any
residual contamination could produce a cross-correlation signal. 
Emission and absorption by our own galaxy 
produce patchy hot spots in the CMB maps
and negative density fluctuations in the galaxy distribution (because
of extinction). In principle, this should therefore result in a
negative cross-correlation, but overcorrecting for the effects of
galactic absorption could also result in a positive signal. 
This possibility have been tested for each of the samples, 
by comparing the cross-correlation to WMAP maps at different frequencies.  
Most analysis use the WMAP Kp0 mask, which
excludes about $30\%$ of sky on the basis of galactic or
extra-galactic (eg radio sources) contamination. 
In all cases the contamination seems smaller than the errors (eg see Fig. 2 in
\scite{Fosalba1}). Moreover, one does not expect this effect to have
any redshift dependence, contrary to the measurements in Fig.
\ref{LCDMgraf}.

Cold dust in distant galaxies, will also produce
patchy hot spots in the CMB maps and positive density fluctuations in
the galaxy distribution (could also be negative 
because of internal extinction). 
The resulting cross-correlation should trace the galaxy-galaxy auto
correlation function, $w_{GG}$, and should therefore have a very
different redshift dependence to the ISW effect. The dotted line
in Fig. \ref{LCDMgraf} shows the predicted shape dependence for
$w_{GG}$ contamination with arbitrary normalization. The shape is
clearly incompatible with the actual cross-correlation measurements.
It is also worth noting how $w_{GG}$ goes quickly to zero at 
$\bar{z} \simeq 0.2$, while the ISW 
cross-correlation remains positive. This is
due to the fact that at these corresponding large scales, $\gsim 40$
Mpc/h, matter-matter correlations $w_{mm}$ effectively 
decays to zero, while $w_{TG}$, which  traces the 
gravitational potential, has a less rapid decay with distance.

\section{Conclusion}

The cross-correlation of CMB anisotropies
with very different galaxy surveys provides consistent detections. Their
combination follows the CCM predictions with a probability of only
$\simeq 3 \times 10^{-5}$ for being a false detection. This provides
new and independent evidence for dark
energy and dark matter, ruling out the EdS model to a high
significance (for any value of $H_0$). Combination with SNIA data
results in strong constraints to  $\Ol = 0.71 \pm 0.13$ and
$\Omm = 0.29 \pm 0.04$.  This in  good agreement with the flat universe
$\Omm+\Ol \simeq 1$ found independently by CMB data \cite{wmap,sdss}.
If we assume a flat universe, we find
$\Ol = 0.70 \pm 0.05$ and $w = -1.02 \pm 0.17$
for a constant dark energy  equation of state.
The data shows, for the first time, 
statistical evidence of a recent slow down in the growth
of structure formation on linear scales, just as expected in 
a flat accelerated universe. The new ISW constraints rely in a totally
different physical effect that previous cosmological 
constraints, providing new light on a dark cosmos.

{\bf Note added in proof:}
After this paper was originally submitted to astro-ph (astro-ph/0407022) 
a related analysis using our data compilation have been published 
by Corasanniti, Giannantonio and Melchiorri  (astro-ph/0504115).

\section*{Acknowledgements} 
{\bf Acknowledgements:} 
We acknowledged support from the Spanish Ministerio de Ciencia i
Tecnologia, project AYA2002-00850 with EC-FEDER funding, and from the
Catalan Departament d'Universitats, Recerca i Societat de la Informaci—.




\begin{thebibliography}{}  

\bibitem[{Afshordi, Loh \& Strauss} <2004>]{Afshordi} Afshordi, N. , Loh, 
 Y., Strauss, M. A., 2004, Phys. Rev. D 69, 083524

\bibitem[{Afshordi} <2004>]{Afshordi2} 
Afshordi,N., 2004 astro-ph/0401166.

\bibitem[{Barriga etal} <2001>]{Subir} 
Barriga, J., Gazta\~naga, E.,Santos,  M.G., Sarkar,S., 2001, MNRAS
324, 977

\bibitem[{Barris et al.} <2004>]{barris} Barris, B. J. 
et al., 2004, ApJ 602, 571.


\bibitem[{Bennett} et al. <2003>]{wmap} 
Bennett C.L. et al., 2003, ApJ Suppl., 148,1 

\bibitem[{Berlind, Naratanan \& Weinberg} <2001>]{berlind} Berlind, A., Naratanan, V., Weinberg, D., 2001, ApJ 549, 688

\bibitem[{Blanchard} <2003>]{Blanchard} 
Blanchard, A., Douspis, M., Rowan-Robinson, M.,
Sarkar, S., 2003, A\&A, 412, 35.

\bibitem[Boldt <1987>]{xray} Boldt, E., 1987, Phys. Rep. 146, 215.

\bibitem[{Boughn \& Crittenden} <2004a>]{Boughn_Nat} Boughn, S., Crittenden, R.
2004, Nature 427, 45 

\bibitem[{Boughn \& Crittenden} <2004b>]{Boughn_A} Boughn, S., Crittenden, R.
2004, astro-ph/0404470 

\bibitem[{Boughn \& Crittenden} <2003>]{BoughnXray}  Boughn, S., Crittenden, R.
2003, astro-ph/0305001 

\bibitem[{Boughn \& Crittenden} <2002>]{BoughnRad} Boughn, S., Crittenden, R. 2002, Phys. Rev. Lett. 88, 021302 


\bibitem[{Condon} etal <1998>]{nvss}  Condon J. J. et al., 1998, AJ 115,
1693

\bibitem[{Crittenden \& Turok} <1996>]{CrTu96}  Crittenden, R. G. Turok,
N., 1996, Phys. Rev. Lett. 76, 575.


\bibitem[{Einseintein \& Hu} <1999>]{EH} D.J. Einseinstein \& W. Hu, 
1998, Astrophys. J. 496, 605

\bibitem[{Fosalba \& Gazta\~naga} <2004>]{Fosalba1} Fosalba, P.
Gazta\~naga, E., 2004, MNRAS 350, 37

\bibitem[{Fosalba, Gazta\~naga \& Castander} <2003>]{Fosalba2} 
Fosalba, P. Gazta\~naga, E., Castander,  F.J., 2003, ApJ 597, L89

\bibitem[{Gazta\~naga \& Baugh} <1998>]{APMPk} 
Gazta\~naga,E. \&  Baugh,  C. M., 1998, MNRAS, 294, 229

\bibitem[{Gazta\~naga \& Lobo} <2001>]{lobo} 
Gazta\~naga, E., \& Lobo, J. A., 2001, ApJ 548, 47


\bibitem[{Jarret} et al. <2000>]{2mass} Jarret, T. H., et al., 2000,
AJ 119, 2498

\bibitem[{Lue \& Starkman} <2004>]{lue} Lue, A., Scoccimarro, R.,
Starkman, G., 2004, Phys. Rev. D, 69,  044005 

\bibitem[{Maddox} et al. <1990>]{APM} Maddox, S. J. ,Efstathiou, G., 
 Sutherland, W. J., Loveday, J., 1990, MNRAS 242, 43P

\bibitem[{Multam\"aki, Gazta\~naga \& Manera} <2003>]{Multamaki1} 
Multam\"aki, T., Gazta\~naga, R., Manera, M., 2003, MNRAS 334, 761

\bibitem[{Nolta} et al. <2004>]{Nolta} Nolta, M. R.  et al. 2004, ApJ 608, 10

\bibitem[{Peacock} et al. <2001>]{peacock} Peacock,  J. A. 
et al., 2001, Nature, 410, 169

\bibitem[{Percival} et al. <2001>]{2df} Percival W.J. et al., 2001, MNRAS, 327, 1297

\bibitem[{Perlmutter} et al. <1999>]{perlmutter} Perlmutter, S. et al., 1999, ApJ 517, 565 

\bibitem[{Pope} et al. <2004>]{pope} Pope,  A. C. et al., 2004,
ApJ, 607, 655

\bibitem[{Riess} et al. <1998>]{riess} Riess A. G. et al., 1998, AJ
 116, 1009 

\bibitem[{Sachs \& Wolfe} <1967>]{ISW} 
 Sachs,R. K., Wolfe,  A. M., 1967, ApJ 469, 437

\bibitem[{Sandvik et al }<2004>]{Tegmark2} Sandvik H.B., Tegmark M., Wang X., andZaldarriaga M., Phys. Rev. D69, 063005

\bibitem[{Scranton et al.} <2003>]{Scranton}  Scranton, R. et al., 2003
astro-ph/0307335.
 
 
\bibitem[{Tegmark} et al. <2004>]{sdss} Tegmark, M. et al., 2004,
ApJ, 606, 70
 






\end{thebibliography}
\end{document}